\documentclass{sf2a-conf2011}
\usepackage{graphicx}
\usepackage[]{natbib}  
\usepackage[cyr]{aeguill}
\usepackage{psfig}
\usepackage{epsfig}

\def\BibTeX{{\rm B\kern-.05em{\sc i\kern-.025em b}\kern-.08em
    T\kern-.1667em\lower.7ex\hbox{E}\kern-.125emX}}
\bibpunct{(}{)}{;}{a}{}{,}  
\newcommand{\kms}{{\,\hbox{km~s$^{-1}$}}}

\newcommand{\spi}{{\it Spitzer}}

\newcommand{\um}{\,\hbox{$\mu$m}}
\newcommand{\taum}{\,\hbox{$\tau_{\rm 9.7}$}}

\begin{document}

\TitreGlobal{SF2A 2011}


\title{Querying for heavily obscured AGN via high 9.7\um\ optical depths: results from the 12\um , GOODS, and FLS Spitzer spectroscopic samples}

\runningtitle{ Compton thick AGN candidates in GOODS and FLS selected via high 9.7 micron optical depths }

\author{K. M. Dasyra}\address{Observatoire de Paris, LERMA (CNRS:UMR8112), 61 Av. de l\'\ Observatoire, F-75014, Paris, France, and \\ 
Laboratoire AIM, Irfu/Service d' Astrophysique, CEA Saclay, Orme des Merisiers, 91191 Gif sur Yvette Cedex, France
}

\author{I. Georgantopoulos}\address{INAF-Osservatorio Astronomico di Bologna, Via Ranzani 1, 40127, Italy, and\\
Institute of Astronomy \& Astrophysics, National Observatory of Athens, Palaia Penteli, 15236, Athens, Greece}

\author{A. Pope}\address{Department of Astronomy, University of Massachusets, Amherst, MA01003, USA}

\author{M. Rovilos}\address{ Max Planck Institut f\"{u}r Extraterrestrische Physik, Giessenbachstra\ss e, 85748, Garching, Germany}

\setcounter{page}{237}

\index{Dasyra, K. M.}
\index{Georgantopoulos, I.}
\index{Pope, A.}
\index{Rovilos, M.}


\maketitle


\begin{abstract}
To optimally identify candidates of the Compton-thick (CT) active galactic nuclei (AGN) that contribute to the unresolved X-ray background in infrared surveys, a tracer of column density is 
desirable in addition to an AGN indicator.  In a recent study, we aimed to test whether the 9.7\um\ silicate absorption feature can be used for this purpose when seen at high optical depths. 
We found that the extreme criterion of optical thickness at 9.7\um\ is efficient in identifying CT objects among local AGN. Having identified six of the nine CT AGN in the 12\um\ sample 
with \spi\ and X-ray spectra, we expanded this analysis at intermediate/high z, using all GOODS and FLS sources with \spi\ and X-ray observations. We found 12 sources with 
\taum $>$1 that host an AGN between 0.8$<$z$<2.7$. Four of them are likely to be CT according to their low X-ray to 6 \um\ luminosity ratio. Surveys with complete coverage in both 
mid-infrared spectra and X-ray data can provide large populations of such sources, as at least 5-9\% of all infrared bright galaxies in the GOODS and FLS samples are \taum $>$1 AGN.
\end{abstract}

\section{Introduction}
  
Observational evidence points out to the existence of a population of active galactic nuclei (AGN) that is yet unidentified due to extreme dust obscuration. The observed space density of black holes in the local Universe cannot be accounted for, unless AGN with column densities of $>$10$^{24}$ cm$^{-2}$ exist \citep{comastri04,merloni08}. At such high column densities, the circumnuclear medium is Compton thick (CT). CT AGN are thought to account for the unresolved cosmic X-ray background \citep{churazov07}, producing 10$-$20\% of the total 30 keV flux  \citep{gilli07,treister09}. Identifying these sources and adding them to the comparison of the black hole accretion rate with the star formation rate history of the Universe is essential, as they are thought to be primarily missing at z$<$1 \citep{gilli07,treister09}, i.e.,  after the bulk of the stellar mass assembly \citep{marconi04,merloni04,gruppioni11}. 
 
The sources missing from the cosmic X-ray background can be sought for in the cosmic background at infrared (IR) or longer wavelengths, as it is in these wavelengths that the radiation absorbed by the dust is re-emitted. Mid-IR excess \citep[e.g.,][]{lacy04,stern05,daddi07,fiore08} and radio excess (Del Moro et al. 2012, in preparation) techniques have been used for this purpose. However, they do not preferentially select type 2 over type 1 AGN. We recently argued that in addition to an AGN tracer,  a high column density indicator can be used to efficiently query for CT objects \citep{georgantopoulos11}. The latter was chosen to be the optical depth of the silicate feature at 9.7\um,  \taum,  when seen in absorption. Even though the 9.7\um\ and X-ray column densities are uncorrelated at moderate optical depths \citep{shi06,wu09}, the bulk of the silicates can still be in a compact, circum-nuclear distribution in front of a bright continuum source \citep{soifer02,tristram07}. They have even been suggested to be located in the AGN torus because their feature is often seen in emission in type-1 AGN and in absorption in type-2 AGN \citep{shi06, hao07}, in agreement with the \citet{antonucci85} unification scheme. We specifically examined how frequently do the local AGN with the most extreme mid-IR obscuration, simply defined as those that are optically thick ($\tau$$>$1) at 9.7\um , have X-ray column densities $>$10$^{24}$ cm$^{-2}$. We then applied this criterion to query for CT candidates among distant AGN samples.

\section{Sample selection}

\subsection{ The 12$\um$ sample}
To quantify the fraction of CT AGN that are identified using the  \taum$>$1 criterion, we used all local 12$\um$-selected Seyferts \citep{rush93} with {\it Spitzer} IRS spectroscopy \citep{wu09}. Re-analysis of these spectra indicated that 11 out of the 103 Seyferts in this sample have  \taum$>$1. Nine of those have X-ray spectra that are necessary to determine their X-ray column densities \citep[e.g.,][]{brightman11,georgantopoulos11}.
 
\subsection{ The GOODS and FLS samples}

\begin{figure}[hb!]
 \centering
 \includegraphics[width=6cm]{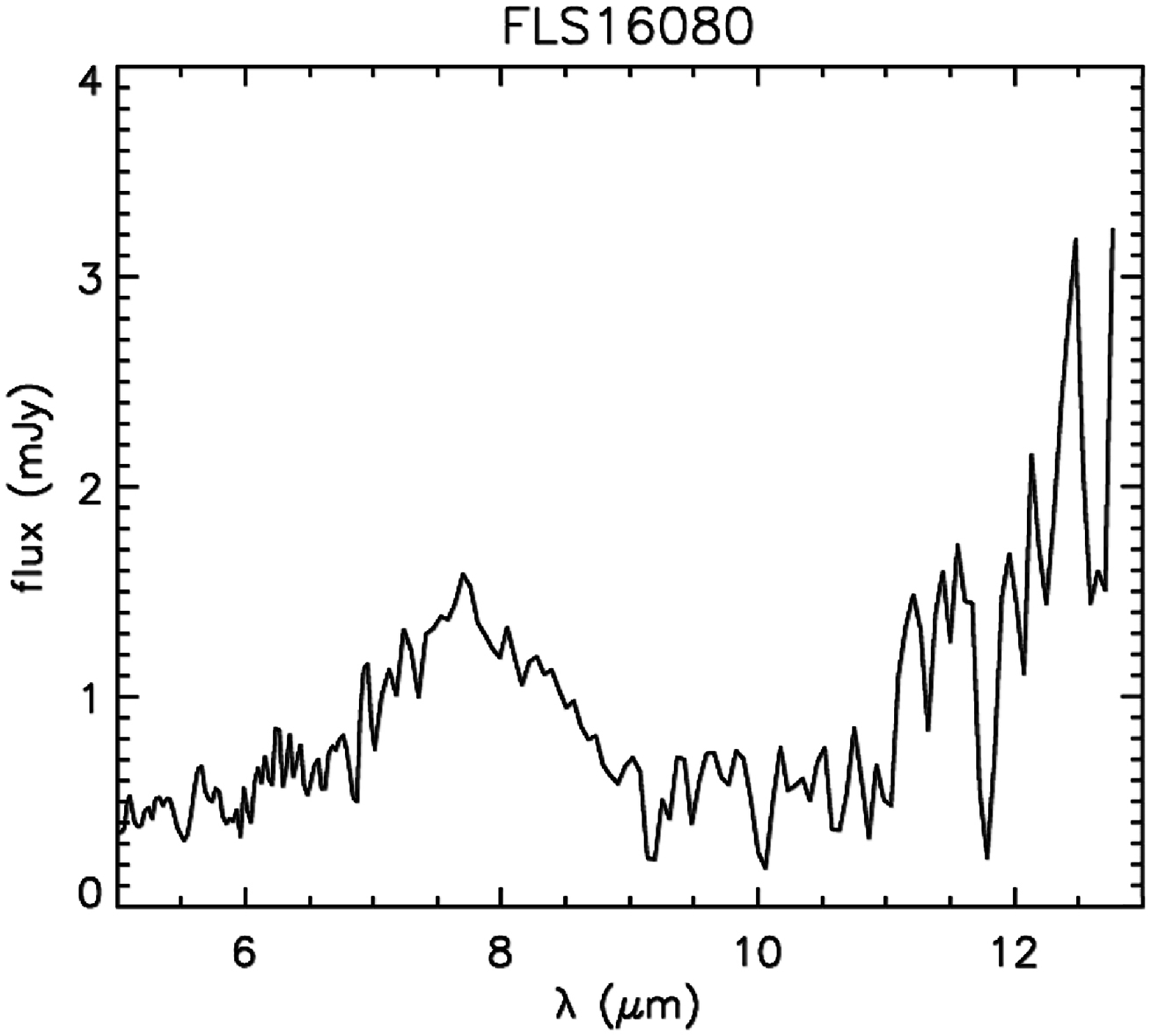}      
 \includegraphics[width=6cm]{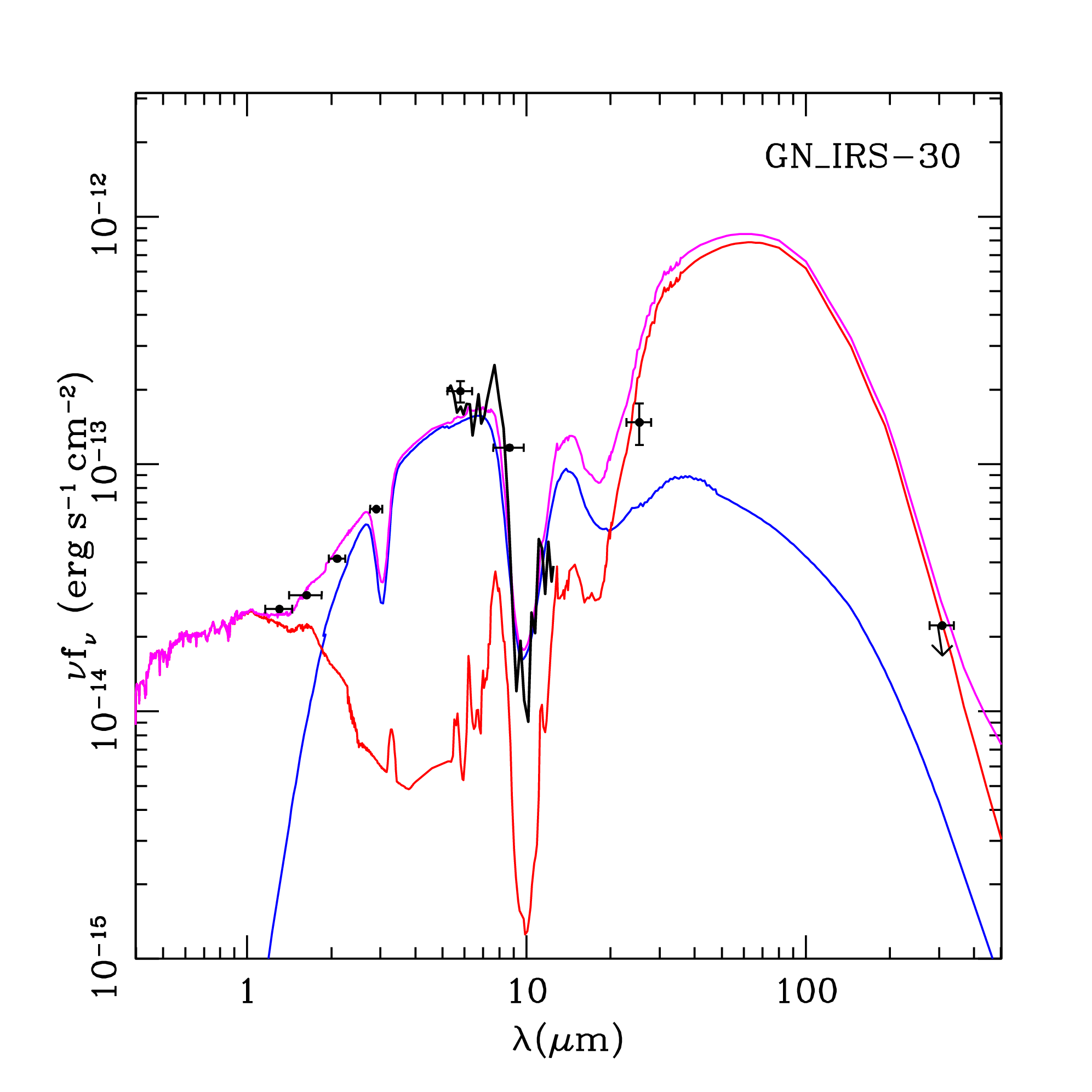}      
  \caption{{\it Left:} The mid-IR spectrum of a distant \taum $>$1 AGN in the FLS \citep{sajina07}. {\it Right:} SED decomposition of a GOODS 
  \taum $>$1 source using an AGN (blue) and a starburst (red) component. The sum of the two is shown in magenta \citep{georgantopoulos11}. }
  \label{dasyra:fig1}
\end{figure}

To identify AGN in the distant Universe that are deeply obscured in the mid-IR (Fig.~\ref{dasyra:fig1}; left panel), we used two surveys with {\it Spitzer} IRS spectroscopy and X-ray data. 
The Great Observatories Origins Deep Survey (GOODS) covering the {\it Chandra} Deep Fields (CDF) North and South, and the First Look Survey (FLS). The two surveys were complementary 
in providing targets. The FLS is a  4\,deg$^2$ shallow survey, whose 220 IRS spectra were flux-limited to a depth of 0.9\,mJy \citep[mainly presented in][]{yan07,sajina07,dasyra09}. 
Its X-ray coverage was nonetheless sparse \citep{bauer10}. On the other hand, the GOODS area  ($\sim$900\,arcmin$^2$ in total) has the deepest mid-IR and X-ray observations available, 
but a sparse (not flux-limited) IRS spectroscopic coverage of 150 sources (Pope et al. 2012, in preparation).  

Overall, 7 \taum $>$1 AGN with X-ray data were identified in GOODS North and South, and 5 in the FLS. The common criterion that we applied to classify distant sources as AGN was the 
lack of strong polycyclic aromatic hydrocarbon (PAH) emission, due either to the dilution of the PAHs into the AGN continuum or to the potential destruction of the PAHs by the AGN 
radiation. It translated to 6.2 (or 11.3) \um\ equivalent widths $<$0.3\um . Several other direct AGN indicators were subsequently found in our sources, including the detection of 
narrow-line-region lines with widths $>$500 \kms , the unambiguous need for an AGN-heated dust component in the spectral energy distribution (SED) fitting (Fig.~\ref{dasyra:fig1}; right panel), 
and the X-ray luminosity L$_X$ values themselves, which exceeded 10$^{42.5}$erg s$^{-1}$ for many sources.

\begin{figure}[ht!]
 \centering
 \includegraphics[width=0.6\textwidth,clip]{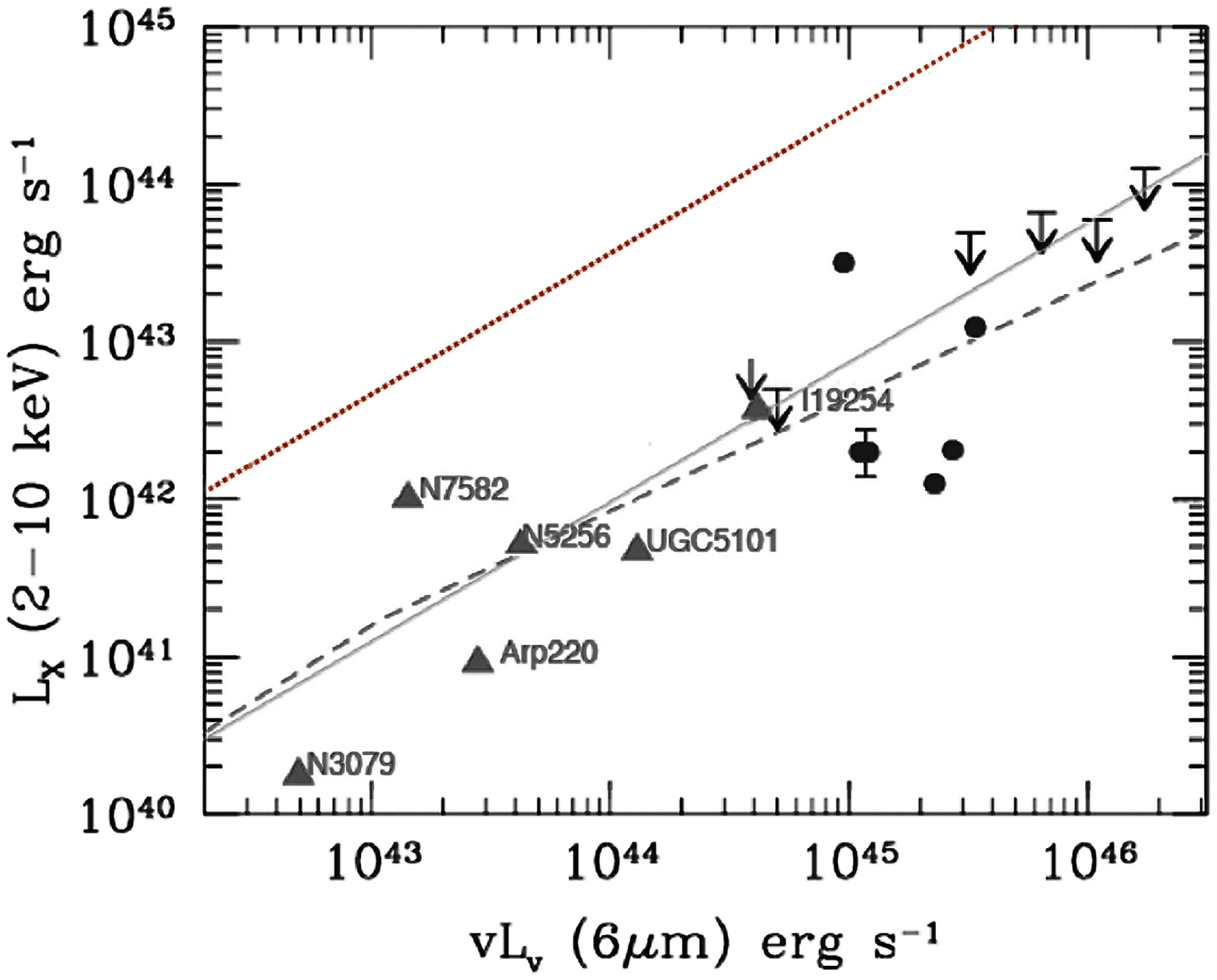}      
  \caption{Observed L$_X$ vs $\nu$L$_\nu(6\um)$ diagram  \citep[adapted from][]{georgantopoulos11}. The solid and dashed lines indicate 
  the area below which CT AGN lie, found for different AGN samples \citep[][respectively]{maiolino07,fiore09} and for a fixed fraction, 0.03, of the 
  intrinsic L$_X$ value being ascribed to the 2-10 keV reflection component luminosity. For comparison, a relation with similar slope to that in \citet{maiolino07}
  that fits the intrinsic (extinction-corrected) X-ray and 6\um\ luminosities of Seyfert 1 and 2 galaxies \citep{lutz04} is shown with a dotted line. Local CT AGN 
  with \taum $>$1 from the 12 \um\ sample are shown as triangles. Distant CT AGN candidates in GOODS and FLS are plotted with circles (or arrows for limits). 
    }
  \label{dasyra:fig2}
\end{figure}

\section{Results}

Nine sources hosting a CT nucleus are known to exist in the sub-sample of 12\um\ Seyferts with both IRS and X-ray spectra. These were identified  
through the detection of (i) a high ($\rm \sim 1\,keV$) equivalent width FeK$\alpha$ line, (ii) a flat X-ray spectrum (with index $\Gamma \sim 1$  
or flatter), which is attributed to reflection from the back side of the torus, or (iii) an absorption turnover at high energies \citep{akylas09}. Six of the
nine sources matching these criteria were found to also have \taum $>$1, suggesting that the identification of CT AGN 
using this \taum\ threshold can be promising despite the different spatial distributions of the media responsible for the X-ray and IR obscuration.

The application of the same \taum\ threshold for the GOODS and FLS samples enabled us to identify 12 CT AGN candidates at 0.8$<$z$<$2.7. Because reliable 
X-ray spectra could not be derived for them, we used the observed L$_X$, integrated over the 2-10 keV range, vs $\nu$L$_\nu(6\um)$ diagram \citep{lutz04} 
to assess which sources could be hosting a CT nucleus. With increasing obscuration, L$_X$ drops and L$_\nu(6\um)$ increases, making the AGN move
from the relation that is appropriate for the intrinsic (extinction-corrected) luminosities (Fig.~\ref{dasyra:fig2}; dotted line) to the bottom-right part of the diagram. 
The boundary of the locus of CT AGN (solid line) is computed using the X-ray and 6\um\ luminosities of the AGN presented in \citet{maiolino07}. Given that this 
boundary depends on the SED and extinction properties of the objects in the chosen sample, we also computed it for the AGN in 
\citet[][shown with a dashed line, and additionally taking into account a luminosity evolution]{fiore09}. 
We placed the 6 local \taum $>$1 CT AGN from the 12$\um$ sample on this diagram, and found that 5 of them are indeed below or close to 
either boundary. For the distant sources, we find that 4 of the 12 candidates are well within the CT range. Several more could be in it, given the proximity
of their upper limits to the boundary. It is thus possible, that the fraction of actual CT AGN among these candidates is high. This remains to be confirmed with 
deep X-ray spectra.

Volume-density wise, our technique could provide a non-negligible fraction of the evasive AGN population. Of the 220 FLS IR-bright galaxies 
with IRS spectra, 20 satisfied both our \taum\ and weak PAH emission criteria. The use of other, direct AGN tracers could make this fraction exceed 
9\%, as CT AGN can also be residing in strong starbursts (with \taum $>$1). In the combined GOODS North and South fields, the fraction of all 150 
IR-bright galaxies with \taum $>$1 was 10\% (15 sources), with 5\% (7) of the sources having weak PAH emission \citep{georgantopoulos11}. 
Potential \taum $>$1 (AGN or starburst) candidates in GOODS were found to correspond to 8$-$16\% of the galaxies with only broad-band {\it Spitzer} 
and {\it Herschel} data \citep{magdis11}. Even in the local Universe, IR-bright galaxies are frequently optically thick at 9.7\um\  \citep{imanishi09}.  

\section{Conclusions}

While not all CT sources have \taum $>$1 due to the clumpy structure of their obscuring medium, the efficiency of identifying them by querying 
for AGN that are optically thick at 9.7\um\ can be high. Six of the nine Seyferts in the local 12\um\ sample that are known to be CT
from their X-ray spectral properties and that have \spi\ IRS spectra satisfied this criterion. In the GOODS and FLS surveys, we found twelve 
sources that are classified as AGN and that have \taum $>$1 , four of which are likely to be CT according to their low X-ray to 6\um\ luminosity 
ratio. While the number of sources presented in this work is limited due to the lack of either X-ray or IR data, the technique has the potential to 
provide large samples of CT AGN candidates. 

\begin{acknowledgements}
This work was supported by the European Community through the Marie Curie Intra-European Fellowships (IEF)  2009-235038 and
2008-235285, which were awarded to K. D. and I. G., respectively, under the 7th Framework Programme (2007-2013).
\end{acknowledgements}


\end{document}